
\def\Ra{\Rightarrow}
\def\LRa{\Leftrightarrow}
\def\lora{\longrightarrow}

\def\pd#1#2{\dfrac{\partial#1}{\partial#2}}

\def\({\left(}
\def\){\right)}
\def\[{\left[}
\def\]{\right]}

\def\a{\alpha}
\def\be{\beta}
\def\g{\gamma}
\def\G{\Gamma}
\def\de{\delta}

\def\la{\lambda}

\def\th{\theta}

\def\vf{\varphi}

\def\S{\Sigma}

\def\coma{\quad ,\quad}
\def\build#1_#2^#3{\mathrel{\mathop{\kern

	0pt#1}\limits_{#2}^{#3}}}

\def\frac#1#2{{#1\over#2}}
\def\dfrac#1#2{{\displaystyle{#1\over#2}}}

\def\sqr#1#2{{\vcenter{\vbox{\hrule height.#2pt

	\hbox{\vrule width.#2pt height#1pt \kern#1pt
            \vrule width.#2pt}
         \hrule height.#2pt}}}}

\def\vspace#1{\crcr\noalign{\vskip#1\relax}}

\def\vs#1{\vspace{#1mm}}

\def\encima#1{\buildrel #1 \over}

\def\section#1{\bigskip\noindent $\P${ \bf #1}}

\def\subsection#1{\medskip\noindent $\bullet${ \sl #1}}

\magnification 1200
\overfullrule 1pt

\hsize=5.3in          \vsize=7.7in
\hoffset=6truemm     \voffset=1truemm
\pretolerance=500     \tolerance=1002    \brokenpenalty=5000
\baselineskip 14 pt    \parskip=6pt plus 2pt  minus 2pt

\font\tengot=eufm10  \font\sevengot=eufm7  \font\fivegot=eufm5

\newfam\gotfam
\textfont\gotfam=\tengot
\scriptfont\gotfam=\sevengot
\scriptscriptfont\gotfam=\fivegot

\def\got{\fam\gotfam\tengot}
\font\sectionart=cmbx10 at 14pt
\font\subsectionart=cmbx10 at 11pt

\centerline{\bf ADMISSIBLE LICHNEROWICZ COORDINATES}

\centerline{\bf  FOR THE SCHWARZSCHILD METRIC}
\medskip
\centerline{J. L. Hern\'andez-Pastora$^\dagger$, J. Mart\'\i
n$^\dagger$, E. Ruiz \footnote{$^\dagger$}{Area de F\'\i sica Te\'orica.
 Edificio Triling\" ue, Universidad de Salamanca. 37008 Salamanca, Spain.
e--mail: jlhp@usal.es} }

\medskip

\vskip  2mm

{\bf Abstract}

Global harmonic coordinates for the complete Schwarzschild metric are found for 
a more general case than that addressed in a previous work by Quan [4]. The
supplementary constant  that appears, in addition to the mass, is related to
the stress quadrupole moment  of a singular energy--momentum tensor.
Similar calculations are also  carried out in q--harmonic coordinates.

\medskip
\noindent{\sectionart 1. - Introduction}
\medskip

Since its advent, the theory of General Relativity has been 
used broadly to handle harmonic coordinates for several purposes. In many cases,
such coordinates are used  for reasons of mathematical comfort  [1],
while other times they have been used with the  purpose of attributing them a
special meaning [2]. In particular, they are  considered to be a natural
generalization of Euclidean Cartesian coordinates. Nevertheless, before
recently  no global system of harmonic and asymptotically  Cartesian coordinates
has been offered for a stellar--like  model; that is, a system  that is well
defined everywhere (interior and exterior) and with a  C$^1$ metric on the
surface of the star (admissible coordinates in the sense of   Lichnerowicz [3]).
One example of this has been provided by  Quan Hui Liu [4]:  the  complete
Schwarzschild  metric (interior and exterior)  with $\mu_0 =  8/9$ ($\mu_0$
being the dimensionless quotient between two times the mass of the  star and its
radius). In this case, it should be noticed that this special value of the 
$\mu_0$  parameter leads to a simplification of the calculations but,
unfortunately, it  also provides a model of a star with divergences in the
pressure at its center.  This programme forces one to introduce two new
constants (in addition to  the mass) into the metric: one of them, 
$Q_{\rm ext}\,$, at the exterior and another, $Q_{\rm int}\,$, in the interior.
These  constants, which are defined with a length dimension in the above paper, 
prove to be proportional to the radius of the star, and according to the author
are physically meaningless.

The aim of the present article is dual: on one hand, we shall develop Quan  Hui
Liu's programme but in a more realistic and general case for the $\mu_0$
parameter ($\mu_0 < 8/9$). The solution of the problem for an arbitrary value
of  this parameter leads to a not well--known Heun differential equation [5] 
instead of the simple hypergeometric equation that appears in the limiting
case   $\mu_0 = 8/9$. On the other hand, by use of the linear
approximation  of vacuum and spherically symmetric Einstein equations we
show that the exterior  constant  $Q_{\rm ext}$  is closely related to the
stress quadrupole moment of  the source.

In Section $2$ the complete  Schwarzschild metric  is written in standard 
Droste coordinates simply to set the notation to be used, and the Darmois
matching  conditions [6] are briefly reviewed by writing the first and the
second   fundamental forms relative to the surface of the star.

In Section $3$ the change of coordinates is found, from standard ones to  
general  asymptotically Cartesian harmonic coordinates,  two  conditions being
fulfilled: the new system of coordinates should preserve the spherical symmetry 
and diagonal structures of the metric (written  in associated polar 
coordinates) and, in addition, the function relating both sets of coordinates 
must be C$^1$ on the surface of the star. These conditions allow us to express 
Quan's exterior and interior  constants in terms of the change functions 
evaluated at the boundary. We then prove that the new coordinates  are 
admissible coordinates in the sense of  Lichnerowicz [3], i.e. the metric is
C$^1$  on the surface.

Section $4$ is devoted to interpreting the exterior constant, $Q_{\rm ext}$. 
To do so, we first write the multipole expansion of the  exterior
Schwarzschild metric in the new set of harmonic coordinates   found earlier
(using the inverse of the radial coordinate, up to order five, as the  parameter
of the series). That expansion has already been shown, although in a  slightly
different way, by one of us [7]. Secondly,  the more  general  solution of the 
vacuum Einstein equations with spherical symmetry in  the linear
approximation is made explicit and hence a  suitable comparison with the
previous  expansion can be established, providing arguments for
understanding the  significance of the new constants appearing in the metric.
Finally, it is  checked that this solution also comes from some singular
 energy momentum tensor whose stress quadrupole moment
proves to be   ``proportional" to the exterior constant of Quan  (which means
that Quan's  constant is the factor multiplying the tensorial expression of that
multipole  moment).

In Section $5$ we briefly discuss the results obtained for the same topic by 
using the so--called q-harmonic coordinates, a variety of harmonic coordinates 
introduced by L. Bel [8]. Most of the contents included here have 
already been obtained by J.M. Aguirregabir\'\i a [9] for the exterior case and 
by  P. Teyssandier [10] for both the exterior and interior cases. Those results,
as  well as some new other ones, are introduced to make a comparison with the 
harmonic scenario and because  until now they have not been published
in their entirety.

The  paper is completed with an Appendix which is devoted to explaining the
Heun differential equation;  this has been included because of  the  rather odd
and at the same time fundamental equation involved. We hope that with this
the reader will find it easy to  understand  Section $3$.

\bigskip
\noindent{\sectionart 2. - Complete model of  the Schwarzschild metric}
\medskip

The interior Schwarzschild metric has the following expression in standard polar 
coordinates $\{t\,,r\,,\th\,,\vf\}$

$$
ds_I^2 = -\[\frac32\,\g^{1/2}(r_0) -
\frac12\,\g^{1/2}(r)\]^2\! dt^2
 + \g^{-1}(r)\,dr^2 + r^2(d\th^2 + \sin^2\!\th\,d\vf^2)
\eqno(1)
$$
where the radial coordinate $r$ is assumed to be restricted to the interval 
$[0,r_0]$,   $r_0$ being the radius of the star. The following notation is  
used:

$$
\g(r) \equiv 1 - \frac{r^2}{L^2} \coma
\frac1{L^2} \equiv \frac13\chi\rho \quad (r_0 < L) \coma \chi \equiv 8\pi
\eqno(2)
$$
$\rho$ being the energy density (constant) of the star and where we are dealing
with  geometrized  units  $(G = c=1)$. As  is known, the pressure of the model
is  given as a function of the radius by the following  expression
$$
p(r) = \rho\frac{\g^{\frac12}(r) - \g^{\frac12}(r_0)}
{3\g^{\frac12}(r_0) - \g^{\frac12}(r)}
\eqno(3)
$$
which provides the following values for the pressure at the surface and the 
center of the  star respectively
$$
\left\{\eqalign{
&p(r_0) = 0  \cr
\vs2
&p(0) \equiv p_c = \rho\frac{1 - \g^{\frac12}(r_0)}
{3\g^{\frac12}(r_0) - 1} \cr
}\right.
\eqno(4)
$$
Since the value $p_c$ of the pressure at the center of the star must be finite, 
it relates the radius of the star $r_0$ and the density parameter $L$ by means 
of the following restriction
$$
3\g^{\frac12}(r_0) - 1 > 0 \quad \Ra \quad \frac{r_0}L <
\frac{2\sqrt{2}}3
\eqno(5)
$$

The exterior Schwarzschild metric is written  in the same set of standard 
polar coordinates as follows ($r\ge r_0$)
$$
ds_E^2 = -\(1-\frac{2m}{r}\)\!dt^2 +
\(1-\frac{2m}{r}\!\)^{\!\!-1}\!\!dr^2 + r^2(d\th^2 +
\sin^2\!\th\,d\vf^2)
\eqno(6)
$$
where $m$ represents the mass of the star, which is related to the density 
$\rho$ and the radius  $r_0$
 by the known expression
$$
m \equiv \frac43\pi r_0^3\,\rho \quad \(\LRa\
\frac{2m}{r_0} = \frac{r_0^2}{L^2} < \frac89\)
\eqno(7)
$$

The above expressions  for the Schwarzschild metric (interior and 
exterior) are such that the Darmois matching conditions [6] are automatically 
fulfilled on the surface of the star ($\S:\ r=r_0$). These conditions --that is,
the continuity of both the first and the second fundamental forms-- will be 
symbolically denoted as follows

$$
\left\{\eqalign{
{\rm I:}&\ \big[h_{ab}\big]_\S = 0 \cr
\vs2
{\rm II:}&\ \big[K_{ab}\big]_\S = 0 \cr
}\right. \coma (a,b,\dots = 0,2,3 = t, \theta , \vf)
\eqno(8)
$$

\noindent In what follows we make explicit the expression of the fundamental 
forms to show the requirements provided by (8).

\subsection{The first fundamental form is}

$$
h_{ab}(p) \equiv g_{\a\be}[x(p)]\,e_{a/}^\a\,e_{b/}^\be
\eqno(9)
$$
$e_{a/}^\a$  being  the tangent vectors to the surface  $\S$ 
$$
e_{a/}^\a(p^b) \equiv
\pd{x^\a}{p^b} \qquad \(\S:\ x^0 = p^0 = t\  ,\  x^2 = p^2 = \th\ ,\ x^3 = p^3 = 
\vf\)
\eqno(10a)
$$
which leads to
$$
(h_{ab}) =
\pmatrix{g_{tt} & 0 & 0 \cr
\vs1
         0 & g_{\th\th} & 0 \cr
\vs1
         0 & 0 & g_{\vf\vf} \cr}
\eqno(10b)
$$
i.e., $g_{tt}$  must be continuous, $g_{rr}$ being  unrestricted (note that
the coordinates satisfy the Euclidean sphere condition everywhere:  
$g_{\vf\vf} = g_{\th\th}\sin^2\!\th = r^2 \sin^2\!\th$). We shall see below that
the continuity  of  $g_{rr}$ is imposed by the continuity of the second
fundamental form.

\subsection{The second fundamental form is}

$$
K_{ab} \equiv -l_\mu
e^\rho_{a/}\nabla_\rho e^\mu_{b/} =
-l_\a\(\pd{e^\a_{a/}}{p^b} +
\G^\a_{\la\mu}e^\la_{a/}e^\mu_{b/}\) = - \G^r_{ab}
\eqno(11)
$$
where the notation $l_\a \equiv \partial_\a(r-r_0)$  has been used  to denote 
the normal vector to the surface  $\S$

$$
(K_{ab}) = \frac12 g^{rr}
\pmatrix{\partial_r g_{tt} & 0 & 0 \cr
\vs1
         0 & \partial_r g_{\th\th} & 0 \cr
\vs1
         0 & 0 & \partial_r g_{\vf\vf} \cr}
\eqno(12)
$$
It may be deduced from (12) that both  $g_{rr}$ and $\partial_r g_{tt}$  must be 
continuous. As a summary,  only the discontinuity of $\partial_r
g_{rr}$ is allowed, which following (1) and (6)  turns out to be

$$
\[\partial_r g_{rr}\]_\S \equiv \partial_r g^E_{rr}(r_0) -
\partial_r g^I_{rr}(r_0) = -3 \frac{r_0}{L^2}\g^{-2}(r_0)
\eqno(13)
$$
It is then obvious that the set of standard polar coordinates 
$\{t\,,r\,,\th\,,\vf\}$ describing the complete Schwarzschild model are not 
admissible in the sense of  Lichnerowicz, since the metric is not C$^1$.

\bigskip
\noindent{\sectionart 3. - Global asymptotically Cartesian harmonic coordinates}
\medskip

The aim of this section is to find a global system of harmonic coordinates for 
the complete Schwarzschild model, which must be asymptotically Cartesian
and  such that the components of the metric are C$^1$ at the boundary
$\S:\,r=r_0$. Let  us consider the generic form of the Schwarzschild metric in
standard coordinates  (1),(6) 

$$
ds^2 = g_{tt}(r)dt^2 + g_{rr}(r)d r^2 + r^2(d\th^2 +
\sin^2\!\th\,d\vf^2)
\eqno(14)
$$
We now perform a change of coordinates 

$$
\{t,r,\th,\vf\}\equiv \{x^\a\} \lora
\{\tilde t,\tilde x,\tilde y,\tilde z\}\equiv \{x^{\tilde\a}\}
\eqno(15)
$$
satisfying the following conditions
$$
\left\{\eqalign{
&\partial_\la\(\sqrt{-g}\,g^{\la\mu}\partial_\mu x^{\tilde\a}\)
=0
\cr
\vs3
&\lim_{\tilde r\to\infty}g_{\tilde\a\tilde\be} = \eta_{\a\be}
\equiv {\rm diag}(-1,+1,+1,+1)\cr
}\right.
\eqno(16)
$$
which represent the harmonicity condition and asymptotic cartesian behaviour 
respectively, and where 
$
\tilde r \equiv \sqrt{\tilde x^2 + \tilde y^2 + \tilde z^2}.
$

The easiest way to solve the problem is by using polar coordinates  $\{t,\tilde 
r,\tilde\th,\tilde\vf\}$  associated with the harmonic ones  $\{x^{\tilde\a}\}$

$$
\left\{\eqalign{
&\tilde x + i\,\tilde y = \tilde r\sin\tilde\th\,e^{i\tilde\vf} \cr
\vs2
&\tilde z = \tilde r\cos\tilde\th
}\right.
\eqno(17)
$$
We see that  in order to preserve the diagonal structure as well as the
structure of the spherical symmetry, the change of coordinates (15) should be as
follows
$$
\tilde t = t \coma \tilde r = f(r) \coma \tilde\th=\th
\coma \tilde\vf = \vf
$$
$$
\(\lim_{r\to\infty} \frac{f(r)}{r} = 1\)
\eqno(18)
$$

Henceforth, equation (16) turns out to be the following second--order linear 
differential equation for the function  $f(r)$:
$$
\frac{d\
}{dr}\(r^2\sqrt{\frac{-g_{tt}}{g_{rr}}}\,\frac{df}{dr}\)-
2\sqrt{-g_{tt}\,g_{rr}}\,f = 0
\eqno(19)
$$
Let us now see  the form of this equation in the interior and at the exterior of
the  star.

{\subsectionart A) The exterior problem}

For the  expressions of the exterior Schwarzschild metric given (6), 
equation  (19) is written as follows
$$
r\(r-2m\)\!f''(r) + 2(r-m)\!f'(r) - 2f(r) = 0
\eqno(20)
$$
whose general solution is as follows [4]
$$
\tilde r = f_{\rm ext}(r) = Q_1(r-m) + Q_{\rm
ext}\,\tilde g(r)
\eqno(21)
$$
where $Q_1$ y $Q_{\rm ext}$ are arbitrary constants and the function 
$\tilde g(r)$ is defined as
$$
\tilde g(r) \equiv (r-m)\log\(1-\frac{2m}{r}\)+2m
\eqno(22)
$$

Looking at the asymptotic condition, we have

$$
f_{\rm ext}(r\to\infty) \sim Q_1r\(1-\frac{m}{r}\) +
Q_{\rm ext}\frac{2m^2}{r} \sim Q_1r
\eqno(23)
$$
and hence the constant $Q_1$ must be equal to one ($Q_1=1$).

{\subsectionart B) The interior problem}

By now introducing  expression (1) from the interior metric into  the
generic  equation (19), we obtain the following differential equation for the
function 
$f(r)$

$$
r^2\g(r)\,f''(r) +
r\[2\g(r)-\frac{r^2}{L^2}\frac{3\g^{\frac12}(r_0)
-2\g^{\frac12}(r)}{3\g^{\frac12}(r_0)
-\g^{\frac12}(r)}\]f'(r) - 2f(r) =0
\eqno(24)
$$
which is at first sight difficult to tackle. Nevertheless, the change of the 
independent variable
$$
r \lora x = 1- \g^{\frac12}(r)\,:\
\left\{\eqalign{
& r \in \big[0\,,r_0\big] \cr
\vs2
& x \in \big[0\,,x_0\equiv 1-\g^{\frac12}(r_0)\big] \cr
}\right.
\eqno(25)
$$
leads to the following equation
$$
 f''(x) + \frac{4x^2-(5+3b)x+3b}{(x-b)x(x-2)}f'(x) -
\frac2{x^2(x-2)^2}f(x) = 0
\eqno(26)
$$
where primes now denote derivatives with respect to the new variable $x$ and
where the  following definition has been used
$$
b \equiv 1-3 \g^{\frac12}(r_0)
\eqno(27)
$$
Let us note here  what restrictions  the upper limit of the 
parameter $\mu_0$ implies:
$$
0 < \mu_0 < \frac89 \quad \Ra\quad\left\{\eqalign{
& 0 < x_0 < \frac23 \cr
\vs2
& -2 < b < 0 \cr
}\right.
\eqno(28)
$$

This new differential equation (26) has  four singular points, all of which are
regular
$$
\{x=b\,,x=0\,, x=2\,,x= \infty\}
\eqno(29)
$$
Now, since we are interested in solutions with good behaviour in the 
 $[0,x_0]$ interval, we look for analytical solutions at the origin as 
a Frobenius series. This procedure leads to a single  analytical series whose
exponent is $E_f = 1/2$. However, this must be constructed by  means of a
rather awkward recursive expression. Accordingly, we have  chosen to reduce 
equation (26) into the  canonical Heun form [5] (see  Appendix), which allows
us to obtain the required solution by using some properties  of that equation.
The final result for the general and analytical solution at  the origin is the
following convergent expression in the  $[0,x_0]$ interval

$$
f_{\rm int}(r) = Q_{\rm int}
\,r\,H\!\(\frac2b,3+\frac1b;4,1,\frac52,1;\frac{x}b\)
\eqno(30)
$$
$Q_{\rm int}$ being an arbitrary constant and where $H(a,q;\a,\be,\g,\de;z)$ is
the so--called Heun series, defined as follows 
$$
H(a,q;\a,\be,\g,\de;z) = 1 + \sum_{n=1}^{\infty}c_n\,z^n
\eqno(31)
$$
with the recursive law ($n=0,1,2,\dots$)

$$
\eqalign{
&a(n+2)(\g+n+1)c_{n+2} \cr
\vs2
&-(n+1)\[\a+\be-\de+n+1+
a(\g+\de+n)+\frac{q}{n+1}\]c_{n+1}\cr
\vs2
&+ \[n(\a+\be+n)+\a\be\]c_n =0 \cr
}
$$

$$
c_0 =1 \coma c_1 = \frac{q}{a\g}
\eqno(32)
$$
which is convergent iff
$$
|a| \ge 1 \coma |z| < 1
\eqno(33)
$$
Note that in our case we have
$$
1<|\frac2b|<\infty \coma |\frac{x}b|< 1 \quad {\rm if}\quad
\mu_0<\frac34
\eqno(34)
$$
i.e., the series is convergent everywhere inside the star only if 
its radius is larger than the Schwarzschild radius by a factor $4/3$.  For 
example, a typical neutron star of $1.4$ solar masses and a radius of $2 \times  
10^4\,m$ gives a parameter 
$$
\mu_0 = \frac{2m}{r_0} = \frac{1.4\times 1.5}{20} =
0.15 < 0.75
\eqno(35)
$$
Thus, $\mu_0 < 3/4$ is  a high upper limit that does not imply restrictions to
a  realistic star model and hence  solution (30) becomes broadly general.

By writing solution (30) as a power series in the parameter $r/L$, we find 
the following expression

$$
\eqalign{
f_{\rm int}(r) =& Q_{\rm
int}\,r\(1+\frac{1+3b}{10\,b}\,\frac{r^2}{L^2} +
\frac{8+19b+45b^2}{280\,b^2}\,\frac{r^4}{L^4}\right. \cr
\vs2
&\phantom{espacio la} +
\left.\frac{152+424b+729b^2+1575b^3}{15120\,b^3}\,\frac{r^6}{L^6}
+\cdots\) \cr }
\eqno(36)
$$

\subsection{Quan's  case}
$$
\mu_0 \equiv \frac{r_0^2}{L^2}=\frac{2m}{r_0}=\frac89\qquad
(\LRa b=0)
\eqno(37)
$$
For this case, solution (30),(36) is strongly divergent and we must return to
the original differential equation (24), taking the parameter $b$ to be  zero ($3
\g^{\frac12}(r_0)=1$). The resulting equation only has three regular singular 
points (one of them at infinity) and hence reduces to a hypergeometric
equation by using the standard procedure (see [4] for the final  result).

{\subsectionart C) Admissible Lichnerowicz coordinates }

We shall prove that it is  possible to set the constants $Q_{\rm int}$ 
and  $Q_{\rm ext}$ in such a way that the components of the metric and their 
derivatives, written in the new set of coordinates, are continuous on the 
surface of the star ($r=r_0$). In order to do so, we only require the function 
$f(r)$ defining the change of coordinates to be C$^1$ at that boundary, i.e.,
$$
\[f\]_\S = \[f'\]_\S = 0 \quad\LRa\quad \left\{\eqalign{
&f_{\rm ext}(r_0) = f_{\rm int}(r_0) \cr
\vs2
&f'_{\rm ext}(r_0) = f'_{\rm int}(r_0) \cr
}\right.
\eqno(38)
$$
which, in agreement with (21-22), (30), can be translated into the following 
relations

$$
\eqalign{
&r_0-m + \tilde g(r_0)\,Q_{\rm ext} =  r_0 \tilde h(r_0)\,Q_{\rm
int}
\cr
\vs2
&1 + \tilde g'(r_0)\,Q_{\rm ext} = \[\tilde h(r_0) +
r_0\tilde h'(r_0)\]Q_{\rm int}\cr
}
\eqno(39)
$$
with
$$
\tilde h(r) \equiv  H\(\frac2b,3+\frac1b;4,1,\frac52,1;\frac{x}b\)
\eqno(40)
$$
The set of equations (39) constitutes an algebraic linear system for the unknown  
variables $Q_{\rm int}$ and $Q_{\rm ext}$,whose solution is
$$
\left\{\eqalign{
Q_{\rm ext} &=
\frac{(r_0-m)[\tilde h(r_0)+r_0\tilde h'(r_0)]-r_0\tilde h(r_0)}
{r_0\tilde h(r_0)\tilde g'(r_0)-[\tilde
h(r_0)+r_0\tilde h'(r_0)]\tilde g(r_0)}
\cr
\vs2
Q_{\rm int} &=
\frac{(r_0-m)\tilde g'(r_0)-\tilde g(r_0)}{r_0
\tilde h(r_0)\tilde g'(r_0)-[\tilde h(r_0)+r_0\tilde
h'(r_0)]\tilde g(r_0)}
\cr
}\right.
\eqno(41)
$$
which is well  defined, since the denominator of both expressions is trivially 
different from zero. We now give the firsts terms of the expansion in power 
series of the $\mu_0$ parameter  for the values of those constants
$$
\left\{\eqalign{
2mQ_{\rm ext} &= r_0\(\frac{12}{35}- \frac{4}{21}\mu_0
-\frac{58}{1155}\mu_0^2 -\frac{136}{5005}\mu_0^3 + \cdots\)\cr
\vs2
Q_{\rm int} &= 1- \frac{3}{4}\mu_0
-\frac{1}{16}\mu_0^2 -\frac{1}{96}\mu_0^3 +
\frac{1439}{134400}\mu_0^4\cdots \cr
}\right.
\eqno(42)
$$
It is worth noticing that these values have nothing to do with those 
obtained by Quan for the case  $\mu_0 = 8/9$, since our results require an upper 
limit for the parameter  $\mu_0 < 3/4$, and hence no comparison can be made.

Let us now check that this choice of constants does indeed lead to a C$^1$
metric, written  in harmonic polar coordinates, at the boundary $r=r_0$. By
making use of the  change (15-18), we have
$$
ds^2 = g_{tt}(r)dt^2 + \frac{g_{rr}(r)}{f'^2(r)}
d\tilde r^2 + r^2(d\th^2 + \sin^2\!\th\,d\vf^2)
\coma \[r=r(\tilde r)\]
\eqno(43)
$$
Obviously, all the components of the metric as well as the derivatives of  
$g_{tt}$, $g_{\th\th}$ and $g_{\vf\vf}$ are continuous. Regarding the 
derivative  of $g_{\tilde r\tilde r}$, this turns out to be
$$
\[\partial_{\tilde r} g_{\tilde r\tilde r}\]_\S =
\frac1{f'}\[\(\frac{g_{rr}}{f'^2}\)'\]_\S =
\frac1{f'^4}\Big(\[\partial_r g_{rr}\]_\S f'(r_0) - 2g_{rr}(r_0)
\[f''\]_\S\Big)
\eqno(44)
$$
However, taking into account the differential equations (20) and (24) for the 
exterior and the interior respectively , we have that
$$
\[f''\]_\S = -\frac32\frac{r_0}{L^2}\g^{-1}(r_0)\,f'(r_0)
\eqno(45)
$$
and therefore, since the discontinuity of $\partial_r g_{rr}$  is given by (13),
we finally have that
$$
\[\partial_{\tilde r} g_{\tilde r\tilde r}\]_\S = 0
\eqno(46)
$$
which is what we wished to prove. Therefore, the harmonic coordinates obtained
with the  constants (41-42) are admissible coordinates in the sense of 
Lichnerowicz.

\bigskip
\noindent{\sectionart 4. - Interpretation of  Quan's exterior constant}
\medskip

{\subsectionart A) Multipole expansion}

As  already mentioned in the Introduction, the purpose of this Section is 
to search for the physical meaning of the constant  $Q_{\rm ext}$ associated
with the choice of global harmonic coordinates given in  the previous Section.
We first carry out a multipole expansion of the exterior  Scwarzschild metric in
those coordinates, i.e., a power series in the inverse of  the radial coordinate
$\tilde r$. This expansion can be taken, up to order five,  from a previous
paper [7], where it was introduced for other purposes. Here we  write
the metric on a slightly different basis that is useful for the following 
calculations.
 $$
ds^2 = T(\tilde r)\,dt^2 + \Big[A(\tilde r)\,
\de_{ij} + B(\tilde r)\,
n_in_j\Big]d\tilde x^i d\tilde x^j
\eqno(47)
$$
with
$$
n^i \equiv \frac{\tilde x^i}{\tilde r} \qquad \(n_i \equiv
\de_{ij}n^j\)
\eqno(48)
$$
$$
T\equiv g_{tt}(r) \coma A\equiv \frac{r^2}{\tilde r^2}
\coma
B\equiv \frac{g_{rr}(r)}{f'^2(r)} -\frac{r^2}{\tilde r^2} \qquad
\Big[r = r(\tilde r)\Big]
\eqno(49)
$$

 By using these definitions and the results from [7], we obtain the following
expansion up  to the order  $1/\tilde r^5$ 

$$
\eqalign{
T &= -1 +2\frac{m}{\tilde r}
-2\frac{m^2}{\tilde r^2}  +2\frac{m^3}{\tilde r^3}
-2\frac{Km +m^4}{\tilde r^4}\cr
\vs3
& \phantom{espacio muy muy largo largo la}+
2\frac{2Km^2+m^5}{\tilde r^5} +\cdots \cr 
\vs4
A &= 1 + 2\frac{m}{\tilde r}
 +\frac{m^2}{\tilde r^2}
+2\frac{K}{\tilde r^3} 
+2\frac{Km}{\tilde r^4}+
\frac65\frac{Km^2}{\tilde r^5} +\cdots \cr
\vs4 
B &=
\frac{m^2}{\tilde r^2}
 +2\frac{-3K+m^3}{\tilde r^3}+2\,\frac{-6Km+m^4}{\tilde
r^4} \cr
\vs3
 & \phantom{espacio muy largo largo}+2\frac{-9Km^2+m^5}{\tilde r^5}
+
\cdots \cr
}
\eqno(50)
$$
where a new constant $K$ is introduced, which is related with  $Q_{\rm ext}$ as 
follows
$$
 K\equiv \frac23 m^3\,Q_{\rm ext}
\eqno(51)
$$

{\subsectionart B) Linear approximation in harmonic coordinates}

As  is well known [11], the Einstein equations can be written in the 
following way
$$
\partial_{\la\mu}({\got g}^{\a\be}{\got g}^{\la\mu} -
{\got g}^{\a\la}{\got g}^{\be\mu}) = 16\pi G(-g)\(T^{\a\be}+
t_L^{\a\be}\)
\eqno(52)
$$
${\got g}^{\a\be}\equiv \sqrt{-g}\,g^{\a\be}$ being the contravariant metric 
density, $g\equiv {\rm det}(g_{\a\be})$ the determinant of the metric, and 
$t_L^{\a\be}$ the  Landau--Lifshitz energy-momentum pseudotensor. One of the 
procedures  used in the past to solve these equations is the so--called 
perturbative postminkowskian algorithm [12]; starting from a system of 
asymptotically Cartesian coordinates, this method consists in looking for a 
solution as a formal power series in the gravitational constant
$G$: 
$$
{\got g}^{\a\be} = \eta^{\a\be} + 
\sum_{n=1}^{\infty}G^n\encima{n}{\got h}{\!}^{\a\be}
\eqno(53)
$$
and taking for the coordinates the harmonicity condition
$$
\partial_\a{\got g}^{\a\be} = 0
\eqno(54)
$$
In particular, this procedure has been of great interest for the study of
certain  aspects of  gravitational radiation [12]. We are now dealing with a
vacuum  stationary scenario --that is, ${\got h}{\!}^{\a\be}$ is independent of
time--  and hence the equations resulting in the linear approximation are as
follows
$$
\left\{\eqalign{
\Delta\,\encima1{\got h}{\!}^{\la\mu} &= 0
\cr
\vspace{1mm}
\partial_i\encima1{\got h}{\!}^{i\mu} &= 0 \cr
}\right.
\eqno(55)
$$
whose general solution can be written in the following way (see, for instance, 
[13])
$$
\encima1{\got h}{\!}^{\la\mu} =\ \encima1{\got h}{\!}^{\la\mu}_{\rm
can}  +\partial^\la w^\mu +\partial^\mu w^\la -
\eta^{\la\mu}\partial_\rho w^\rho\coma (\,\Delta\, w_\la=0)
\eqno(56)
$$
where the canonical part $\encima1{\got h}{\!}^{\la\mu}_{\rm
can}$ is given by
$$
\left\{\eqalign{
\encima1{\got h}{\!}^{00}_{\rm can} &= -4\sum_{l=0}^{\infty}
\frac1{r^{l+1}}\sum_{m=-l}^{+l}M^m_l\,Y^m_l(\theta,\vf) \cr
\vs2
\encima1{\got h}{\!}^{0j}_{\rm can} &= +4\sum_{l=1}^{\infty}
\frac1{r^{l+1}}\sum_{m=-l}^{+l}S^m_l\,Y^{m,j}_{l,l}(\theta,\vf)
\cr
\vs2
\encima1{\got h}{\!}^{jk}_{\rm can} &= 0
}\right.
\eqno(57)
$$
$M^m_l$ and  $S^m_l$  being  the Geroch--Hansen static and dynamical multipole
moments [14] (up to a numerical constant factor), and where the usual 
notation has been used for spherical harmonics. Regarding  the ``gauge" part of
the  solution [terms involving the function $w^{\rho}$ in (56)], we have the
following  expressions
$$ 
w^0 = \sum_{l=0}^{\infty}\frac1{r^{l+1}}
\sum_{m=-l}^{+l}C^m_lY^m_l(\th,\vf)
\eqno(58a)
$$
$$
\eqalignno{
w^k = &\ \sum_{l=1}^{\infty}
\frac1{r^l}\sum_{m=-l}^{+l}H_l^{m,k}\, Y_l^m(\th,\vf) 
+\sum_{l=1}^{\infty}\frac1{r^{l+1}}
\sum_{m=-l}^{+l}J_l^m \, Y_{l,l}^{m,k}(\th,\vf) \cr
\vs3
&-\sum_{l=0}^\infty\frac{l+1}{r^{l+2}}
\sum_{m=-l}^{+l}K_l^m\,  Y_l^m(\th,\vf)\,n^k 
+\sum_{l=0}^\infty\frac1{r^{l+1}}
\sum_{m=-l}^{+l}K_l^m\, \partial^k Y_l^m(\th,\vf) &(58b)\cr
}
$$
where $C^m_l$, $H_l^{m,k}$, $J_l^m$ y $K_l^m$ are new constants (scalar or 
vectorial) whose meanings have not yet been discussed.

The general solution of  equations (55) with spherical symmetry is obtained 
from the previous expressions (57-58) by restricting ourselves to 
monopolar  terms, i.e, by taking into account only  $l=0$ in the above series:
$$
\left\{\eqalign{
\encima1{\got h}{\!}^{00}_{\rm can} &= -4\frac{m}{r} \qquad
\(m\equiv M^0_0\)\cr
\vs1
\encima1{\got h}{\!}^{0j}_{\rm can} &= 0 \cr
\vs1
\encima1{\got h}{\!}^{ij}_{\rm can} &= 0 \cr
}\right.
\coma
\left\{\eqalign{
w^0 &= \frac{C}{r} \qquad \(C\equiv C^0_0\)\cr
\vs2
w^j &= -\frac{K}{r^2}n^j \qquad \(K\equiv K^0_0\)\cr
}\right.
\eqno(59)
$$
In agreement with (56), this expressions lead to the following solution
$$
\left\{\eqalign{
\encima1{\got h}{\!}^{00} &= -4\frac{m}{r} \cr
\vs1
\encima1{\got h}{\!}^{0j} &= \partial^j w^0 \cr
\vs1
\encima1{\got h}{\!}^{ij} &= -2\frac{K}{r^3}\de^{ij} +
6\frac{K}{r^3}n^i n^j \cr
}\right.
\eqno(60)
$$
Finally, the linear order of the metric is obtained by using  the known 
relation
$$
\encima1{g}_{\la\mu} = - \encima1{\got h}_{\la\mu} + 
\,\,\frac12\encima1{\got h}\eta_{\la\mu} \qquad \(\encima1{\got h}
\equiv \eta_{\la\mu}\encima1{\got h}^{\la\mu}\)
\eqno(61)
$$
which explicitly turns out to be 
$$
\left\{\eqalign{
\encima1g_{00} &= 2\frac{m}{r} \cr
\vs2
\encima1g_{0j} &= \partial_j w^0 = 0 \coma (C=0) \cr
\vs2
\encima1g_{jk} &= \(2\frac{m}{r} + 2\frac{K}{r^3}\)\delta_{jk}
-6\frac{K}{r^3}n_j n_k\cr
}\right.
\eqno(62)
$$
where we have taken $C=0$ since $\encima1g_{0j}$
is a gradient and can therefore be omitted (static condition).

Let us notice that, as it should,  metric (62) contains all the linear 
terms of the Schwarzschild metric written in harmonic coordinates (47),(50) 
shown in the previous Section. This detail, which might appear 
insignificant, turns out to be the relevant  point for understanding the meaning
of  the constant $K$, as we shall now see. 

{\subsectionart C) A spherically symmetric  ``singular" source}

A significant feature of the metric (62) is that it proves to be  a solution 
of the linearized Einstein equations with the following singular energy-
momentum tensor (order zero) on the right hand side of the equations,
$$
\left\{\eqalign{
\encima0{T}{\!}^{00} &=  m\, \de(\vec x)  \cr
\vs2
\encima0{T}{\!}^{0j} &= 0 \cr
\vs2
\encima0{T}{\!}^{ij} &= \frac12 K\[\de^{ij}\de^{kl}-
\de^{i(k}\de^{l)j}\]
\partial_{kl}\de(\vec x) \cr
}\right.
\eqno(63)
$$
Thus, the equations can be written as follows:
$$
\left\{\eqalign{
\Delta\encima1{\got h}{\!}^{00} &= 16\pi m\,\de(\vec x) \cr
\vs3
\Delta\encima1{\got h}{\!}^{ij} &=  8\pi K\[\de^{ij}
\Delta\de(\vec x) -
\partial^{ij}\de(\vec x)\]
\cr }\right.
\eqno(64)
$$
and we can solve them by using the Poisson integrals to give
$$
\left\{\eqalign{
\encima1{\got h}{\!}^{00}(\vec x) &= -4m\int 
\frac{\de(\vec y)}{|\vec x-\vec y|}d^3\vec y = -4\frac{m}{r} \cr
\vs4\encima1{\got h}{\!}^{ij}(\vec x) &=  + 2K
\int\frac{\partial^{ij}\de(\vec y)}{|\vec x-\vec y|}d^3\vec y =
-2\frac{K}{r^3}\de^{ij} + 6\frac{K}{r^3}n^i n^j\cr
}\right.
\eqno(65)
$$
which is merely the metric density obtained before (60).

Let us now calculate  the {\it stress quadrupolar moment} of this energy-
momentum tensor (63)
$$
\eqalign{
\int x^p x^q \encima0{T}{\!}^{ij}\,d^3\vec x &=
\frac12K\[\de^{ij}\de^{kl}-
\de^{i(k}\de^{l)j}\]\int x^p x^q\partial_{kl}\de(\vec x)
\,d^3\vec x \cr
\vs2
&= K\[\de^{ij}\de^{pq}- \de^{i(p}\de^{q)j}\] \cr
}
\eqno(66)
$$
We see that the constant $K$, and therefore  $Q_{\rm ext}$, are  related to the
stress quadrupolar moment of the source. In our  opinion, this is simply a first
argument to understand the role played by this kind  of  constant, which are
not present when the  Thorne gauge is used for the analysis of standard
multipole moments. In the future we hope to develop a coherent  theory able to
completely justify the interpretation made.

\bigskip 
\noindent{\sectionart 5. -  Global asymptotically Cartesian q-harmonic 
coordinates}
\medskip

In Section $3$ we  looked for a global system of harmonic coordinates that
would be asymptotically Cartesian and admissible in the sense on
Lichnerowicz. This  issue can also be solved for a system of the so--called
q-harmonic coordinates.  These coordinates were introduced by Bel [8] to analyze
reference frames   in  General Relativity as  congruences of time-like curves.
Although the following  results are (almost all) from other authors, we shall
show them  briefly because they have not been published by those authors.

Q-harmonic coordinates are only meaningful when associated with the also
so--called  q-harmonic congruences [8]. Since the Schwarzschild metric admits as
a q-harmonic congruence the  Killing time congruence, we restrict ourselves to
recalling the definition of q-harmonic coordinates for this case. The change of 
coordinates from the standard ones would be as follows
$$
\{r,\th,\vf\}\equiv \{x^i\} \lora
\{\bar x,\bar y,\bar z\}\equiv \{x^{\bar i}\}
\eqno(67)
$$
with the q-harmonicity condition
$$ 
\left\{\eqalign{
&\partial_i\(\sqrt{\hat g}\,g^{ij}\partial_j x^{\bar k}\) =0 \cr
\vs3
&\lim_{\bar r\to\infty}g_{\,\bar i\,\bar j} = \de_{ij} \cr
}\right.
\eqno(68)
$$
where $\bar r \equiv \sqrt{\bar x^2 + \bar y^2 + \bar z^2}\,$, and $\hat g$ 
represents the determinant of the spatial metric of the quotient space defined 
by the Killing congruence.

As  in the harmonic case, associated polar coordinates are defined by
$$
\left\{\eqalign{
&\bar x + i  \bar y = \bar r\sin\bar\th\,e^{i\bar\vf} \cr
\vs2
&\bar z = \tilde r\cos\bar\th
}\right.
\eqno(69)
$$
$$
 \bar r = f(r) \coma \bar\th=\th  \coma \bar\vf = \vf
\eqno(70)
$$
$$
\(\lim_{r\to\infty} \frac{f(r)}{r} = 1\)
$$
and this leads to the following differential equation
$$
\frac{d\
}{dr}\(\frac{r^2}{\sqrt{g_{rr}}}\frac{df}{dr}\)-2\sqrt{g_{rr}}\,f
= 0
\eqno(71)
$$

{\subsectionart A) The exterior problem (q--harmonic coordinates)}

In this case the previous differential equation (71) is written  as
follows
$$
r\(r-2m\)\!f''(r) + 2(r-\frac32m)\!f'(r) - 2f(r) = 0
\eqno(72)
$$
whose general solution was already obtained by Aguirregabir\'\i a [9]
$$
\bar r = f_{\rm ext}(r) = J_1\(r-\frac32m\) + J_2\,\bar g(r)
\eqno(73)
$$
where $J_1$ and $J_2$ are constants, and the function
$\bar g(r)$ is defined by
$$
\bar g(r) \equiv \(r-\frac12m\)\sqrt{1-\frac{2m}{r}}
\eqno(74)
$$
with the following asymptotic behavior
$$
\bar g(r) = r\(1-\frac32\frac{m}{r} - \frac14\frac{m^3}{r^3} +
\cdots \)
\eqno(75)
$$
Thus, according to (73) we have
$$
J_1 + J_2 =1
\eqno(76)
$$

{\subsectionart B) The interior problem (q--harmonic coordinates)}

For this case, the differential equation (71) has the following form
$$
r^2\(1-\frac{r^2}{L^2}\)\!f''(r) +
r\(2-\frac{3r^2}{L^2}\)\!f'(r) - 2f(r) = 0
\eqno(77)
$$
whose unique analytical solution at the origin has already been obtained by
Teyssandier [10]
$$
\eqalign{
f_{\rm int}(r) &=
P\,\frac{r}{L}\,F\(\frac12,\frac32,\frac52;\frac{r^2}{L^2}\)\cr
\vs3
&= P\,\frac{r}{L}\(1+\frac3{10}\frac{r^2}{L^2} +
\frac9{56}\frac{r^4}{L^4} + \frac{5}{48}\frac{r^6}{L^6}
+ \cdots \)
\cr }
\eqno(78)
$$
$P$ being an arbitrary constant, and where $F(a,b,c,;x)$ represents the usual
hypergeometric function whose first terms of its  power  expansion  are shown
for clarity of expression.

{\subsectionart C) Continuity at the boundary $\S$}

As  in the harmonic case, the arbitrary constants $P$ and  $J_2$ (or 
$J_1$) can be fixed by imposing that the function  $f(r)$ be C$^1$ on the 
surface of the star, i.e.,
$$
\[f\]_\S = \[f'\]_\S = 0 :\ \left\{\eqalign{
&\(r_0-\frac32m\)J_1 + \bar g(r_0)\,J_2 =  r_0
\bar h(r_0)\,\frac{P}{L}
\cr
\vs2
&J_1 + \bar g'(r_0)\,J_2 = \[\bar h(r_0) +
r_0\bar h'(r_0)\]\frac{P}{L}
\cr }\right.
\eqno(79)
$$
with
$$
\bar h(r) \equiv  F\(\frac12,\frac32,\frac52,\frac{r^2}{L^2}\)
\eqno(80)
$$
These conditions lead to the following expressions for the constants
$$
\left\{\eqalign{
J_2 &=
\frac{(r_0-3m/2)\[\,\bar h(r_0)+r_0\bar h'(r_0)\]-r_0\bar h(r_0)}
{r_0\bar h(r_0)[\bar g'(r_0)-1]-\[\,\bar
h(r_0)+r_0\bar h'(r_0)\]\![\bar g(r_0)-(r_0-3m/2)]}  \cr
\vs3
\frac{P}{L} &=
\frac{(r_0-3m/2) \bar g'(r_0)-\bar g(r_0)}
{r_0\bar h(r_0)[\bar g'(r_0)-1]-\[\,\bar
h(r_0)+r_0\bar h'(r_0)\]\![\bar g(r_0)-(r_0-3m/2)]}\cr
}\right.
\eqno(81)
$$
whose expansion in power series of the parameter $\mu_0\equiv 2m/r_0 =
r_0^2/L^2$ is given by the following expressions
$$
\left\{\eqalign{
&4m^2J_2 = r_0^2\[\frac{-8}5 +
\frac{72}{35}\mu_0-\frac{29}{1050}\mu_0^2 -
\frac{8}{825}\mu_0^3 +
O(\mu_0^4)\] \cr
\vs3
&\frac{P}{L} = 1-\mu_0 +\frac{9}{80} \mu_0^2+\frac{1}{280} \mu_0^3+ O(\mu_0^4) 
\cr
}\right.
\eqno(82)
$$
Finally, it should be mentioned that  Teyssandier [10] has proved, in a similar 
way to that used here with the harmonic coordinates, that these values of the
constants  provide a C$^1$ metric on the surface  $\S$, i.e.,  the
q-harmonic  coordinates obtained  are also admissible coordinates in the sense
of Lichnerowicz.

\bigskip
\noindent{\sectionart 6. - Conclusions}
\medskip
We have presented a global solution to Einstein equations for static spherically
symmetric perfect fluid with homogeneous energy density distribution. At the
outside of matter distribution we have, as it should be, the Schwarzschild
space--time, whereas inside, our solution coincides with the well--known
Schwarzschild solution. The novelty of our approach being that the solution is
written globally in harmonic and q--harmonic coordinates. At the boundary
surface the Lichnerowicz junction conditions are satisfied, which implies that,
both, the metric and its first derivatives are continuous across the boundary.
This, in  turn, implies that the exterior space--time contains an additional
paramenter, besides the mass. This parameter being related to the stress
quadrupolar moment [see (66)]. Obviously this new parameter may be
eliminated by means of a coordinate transformation. All the above brings out two
facts: on one hand, the Schwarzschild space--time may be produced by a variety
of sources (which is well-known) and on the other hand, when the solution is
written globally, using Lichnerowicz junction conditions, it contains an
additional parameter reflecting properties of the source, other than the mass.

\bigskip
\noindent{\sectionart Appendix :}
\medskip
\noindent{\subsectionart Heun's equation}

{\bf A)} The so--called Heun equation is the following second--order linear 
differential equation

$$
\eqalign{
&x(x-1)(x- a) F''(x) +\{(\a+\be +1)x^2 \cr
\vs2
&-[\a+\be +1+ a(\g+\de)-\de]x+  a\g\} F'(x) + (\a\be x- q)F(x) =0
\cr }
\eqno(A1)
$$
where $(a,q;\a,\be,\g,\de)$ are numerical constants. This equation has the 
following singular points,  all of which are  regular
$$
\{x=0\,,x=1\,, x=a\,,x= \infty\}
\eqno(A2)
$$
We restrict ourselves to the point $x=0$, which is the one of interest in the 
main text, and one readily sees that the Frobenius series type solutions
have  exponents $(0\,,1-\g)$.

For the zero exponent, the solution is 
$$
F(x) = H(a,q;\a,\be,\g,\de;x) \coma (\g \neq 0,-1,-2,\dots)
\eqno(A3)
$$
$H$ being the Heun series (31) appearing  in the main text, which is convergent
if  
$|a|\ge 1$ and with a radius of convergence $|x|<1$.

For the exponent $1-\g$, the solution is
$$
F(x) = x^{1-\g}H(a,q_1;\a_1,\be_1,\g_1,\de;x) \coma (\g \neq
1,2,3,\dots)
\eqno(A4)
$$
with

$$
\left\{\eqalign{
&q_1 \equiv q+(1-\g)(\a+\be+1-\g-\de+a\de) \cr
\vs2
&\a_1 \equiv \a+1-\g \coma \be_1 \equiv \be+1-\g \coma
\g_1 \equiv 2-\g \cr
}\right.
\eqno(A5)
$$

When the previous series are not convergent, i.e., $|a|<1$, we
perform the change of variable

$$
x \lora \bar x = \bar a x \coma \(\bar a \equiv \frac1a\)
\eqno(A6)
$$
and this leads to the following solution

$$
\left\{\eqalign{
&E_f = 0\,:\quad F(x) = H(\bar a,\bar q;\a,\be,\g,\bar\de;\bar x)
\cr
\vs2
&E_f = 1-\g\,:\quad F(x) = x^{1-\g}   H(\bar a,\bar q_1
;\a_1,\be_1,\g_1,\bar\de;\bar x) \cr
}\right.
\eqno(A7)
$$
with

$$
\bar q \equiv \bar a\, q \coma \bar \de \equiv \a+\be+1-\g-\de \coma
\bar q_1 \equiv \bar a\, q_1
\eqno(A8)
$$

{\bf B)} The differential equation (26) appearing in the main text can be
reduced to  the Heun canonical form by means of the following change of
function 
$$
f \lora F\,:\quad f(x) = x^k(x-2)^l(x-b)^m\, F(x)
\eqno(A9)
$$

A rather awkward but straightforward calculation shows that there are four 
sets of suitable values for the exponents $(k\,,l\,,m)$ and consequently for the 
corresponding parameters $(a\,,q\,;\a\,,\be\,,\g\,,\de)$.  Since we are looking 
for an analytical solution of  equation (A1) in the neighbourghood of the 
origin ($x=0$), we obtain  eight  possibilities, provided by  the two Frobenius
exponents  $(0\,,1-\g)$ of the Heun equation at this point.  Taking into account
that the only correct Frobenius exponent from the starting  equation (26) is 
$1/2$, we finally have only four possibilities. It can be  checked that,
naturally, all these four possibilities lead to the same  analytical  solution
at  $x=0$,  one of them being defined by the following set of parameters
$$
E_f=0 \quad \left\{\eqalign{
&k=\frac12\coma l=\frac12 \coma m=0 \cr
\vs2
&a=\frac{b}2 \ ,\ q=3\frac{b}2 +\frac12 \ ;\  \a =4
\ ,\  \be =1 \ ,\  \g=\frac52 \ ,\ \de = \frac52 \cr
}\right.
\eqno(A10)
$$
Now, since  $|a|< 1$, we must resort to formulae (A6), (A8), which provide the
following values for the parameters
$$
\bar a=\frac2b \coma \bar q=3+\frac1b \coma \a=4 \coma \be =1
\coma \g = \frac52 \coma \bar\de = 1
\eqno(A11)
$$
which construct the solution shown in the main text (36).

\vfill\eject

\noindent{\sectionart Acknowledgments}
\medskip
This work was supported by a Research Grant from the ``Ministerio de Ciencia y
Tecnolog\'{\i}a" of Spain, number BFM2000--1322, and the authors 
thank J.M.M. Senovilla for helpful comments on the Heun differential
equation.

\noindent{\sectionart References}
\medskip

\noindent [1] {\it See for instance},  A. Lichnerowicz,  {\it Relativistic
Hydrodynamics and Magnetohydrodynamics\/}, Benjamin Inc. (1967)

C. W. Misner, K. S. Thorne and J. A. Wheeler, \  {\it Gravitation}, W. H. 
Freeman \& Co. San Francisco. (1970).

\noindent [2] V. A. Fock,  \ {\it The Theory of Space, Time and Gravitation}, 
$2^o$ ed. revision Mc Millan. New York, translated by N. Kemmer. (1964)

 Ll. Bel, J. Martin and A. Molina,  \ {\it Journal of the Physical Society of 
Japan}, {\bf 63}, 4350--4363 (1994).

Recommendations I, II, III by  the XXI General Assembly of the International 
Astronomic Union.

\noindent [3] A. Lichnerowicz,  \  {\it Th\'eories Relativistes
de la Gravitation et de l'\'Electromagn\'etisme\/}, Masson \& Cie. (1955)

\noindent [4] Quan-Hui Liu,  \ {\it Journal of Math. Phys.}, {\bf 39}, 6086--
6090. (1998).

L. A. Gegerly, \ {\it Journal of Math. Phys.\/}, {\bf 40}, 4177--4178. (1999).

Quan-Hui Liu, \ {\it Journal of Math. Phys.\/}, {\bf 40}, 4179. (1999).

\noindent [5] K. Heun, \ {\it Math Annalen\/} {\bf 33}, 161--179. (1889).

E. Kamke,  \ {\it Differentialgleichungen. L\"osungsmethoden und l\"osungen\/}, 
Chelsea Publishing Co. New York. 485--486. (1971).

E. L. Ince,  \ {\it Ordinary differential equations\/}, Dover Publications Inc.. 
New York. 394. (1956).

\noindent [6] {\it See for instance}, C. W. Misner, K. S. Thorne and J. A. 
Wheeler, \  {\it Gravitation}, W. H. Freeman \& Co. San Francisco. (1970).

\noindent [7] J. M. Aguirregabir\'\i a, Ll. Bel, J. Mart\'\i{}n, A. Molina
 and E. Ruiz, {\it Gen. Rel. Grav.\/}, {\bf 33}, 1809 (2001). Preprint
gr-qc/ 0104019.

\noindent [8] Ll. Bel  and J. Llosa, \ {\it Gen. Rel. and Grav.}, {\bf 27}, 
1089--1110. (1995).

\noindent [9] J. M. Aguirregabir\'\i a, \ {\it private communication}. See also 
[7].

\noindent [10] P. Teyssandier, \  Preprint from  Laboratoire de
Gravitation et Cosmologie Relativistes. CNRS/ESA 7065 (France)

\noindent [11] {\it See for instance},  L. Landau et E. Lifchitz, \ {\it
Th\'eorie des Champs\/}, Editions MIR (Moscou 1970).

\noindent [12] K. S. Thorne, \  {\it Rev. Mod. Phys.\/}, {\bf 52}, 299--339 
(1980)

Ll. Bel, T. Damour, N. Deruelle, J. Ib\'a\~nez  and J. Mart\'\i n, \  {\it Gen.
Rel. Grav.\/}, {\bf 13}, 963--1004 (1981).

L. Blanchet  and T. Damour, \  {\it Phil. Trans. R. Soc.
Lond.\/} {\bf A 320}, 379--430 (1986).

\noindent [13] K. S. Thorne, \ {\it Rev. Mod. Phys.}
{\bf 52}, 299. (1980).

\noindent [14] R. Geroch,  \ {\it J. Math. Phys.},
{\bf 11}, 1955. (1970).

R. Geroch, \ {\it J. Math. Phys.},
{\bf 11}, 2580. (1970).

R. Geroch,  \ {\it J. Math. Phys.},
{\bf 12}, 918. (1970)

R. O. Hansen,   \ {\it J. Math. Phys.},
{\bf 15}, 46. (1974).

\bye